# Benchmarking Transparent Conductors

**Amit Cohen, Lior Kornblum**

Andrew and Erna Viterbi Department of Electrical and Computer Engineering,

Technion—Israel Institute of Technology, Haifa 32000-03, Israel

amit.cohen@campus.technion.ac.il , liork@technion.ac.il

## Abstract

Transparent conducting oxides (TCOs) are central to optoelectronic technologies, yet their design is often guided by popular figures of merit that are disconnected from the electrical requirement of actual devices. As a result, widely used metrics guide material design under conditions that can be impractical for devices. Here, we introduce a benchmarking framework to guide TCO development, in which transparent conductors are evaluated at fixed, application-relevant sheet resistance ($R_S$). The resulting metric, $T_{app}(R_S)$, anchors comparison to device requirements, asking instead: *What optical transparency can be obtained at the sheet resistance required by a given application?* This approach provides a directly interpretable measure of performance, enabling materials to be benchmarked in terms of absolute transparency gains at a specified $R_S$. Applied to representative conventional and emerging TCOs, the framework defines the sheet-resistance landscape relevant to each application and maps how different materials perform within it. In doing so, it provides an application-rooted guide to material development and selection. More broadly, this approach establishes a general strategy for evaluating materials under fixed operational constraints, bridging the gap between materials design and device integration.

## Main

Transparent conducting oxides (TCOs) are central to modern optoelectronic technologies, where they function as a front electrode that transmits light and collects or injects electrical currents.[1] Improving TCOs remains a materials-design challenge due to the competing requirements of low optical absorption and low electrical resistivity, with direct implications for photovoltaics, displays and light-emitting devices.[2–5] To evaluate TCO performance and guide material development, figures of merit (FOMs) that combine optical transmittance (T) and sheet resistance ($R_s$) into a single metric are widely used, most notably the Haacke FOM[6] ($\phi = T^{10}/R_S$).

However, such approaches implicitly treat sheet resistance as a free variable, often misidentifying optimal performance under conditions that are not realistic for practical electrodes. Alternatively, TCO performance can be assessed within complete devices, but such evaluations depend strongly on system-specific details and therefore offer limited guidance for material development beyond a specific architecture.[7–10] This leaves an application gap between materials benchmarking and device engineering: TCOs lack a general, materials-level framework that nevertheless acknowledges the electrical constraints imposed by practical applications.[11] Here, we introduce a framework that anchors evaluation to application-relevant sheet resistance and uses spectrally weighted transmittance to quantify optical performance at device-realistic operating points. This approach establishes a general strategy for assessing materials under fixed operational constraints, enabling material development to be guided by the sheet-resistance regimes required by target applications.

We therefore introduce a benchmarking framework in which transmittance is evaluated at application-specific target sheet resistances, allowing TCOs to be compared directly under device-relevant equal-$R_s$ conditions. This framework is named here as benchmarked electrical sheet-resistance transmittance (BEST). Within the BEST framework, we define an application-specific performance metric,

$$T_{app} = \frac{\int T(\lambda, R_S)\, W(\lambda)\, d\lambda}{\int W(\lambda)\, d\lambda} \qquad (1)$$

Where $T_{app}$ is the spectrally-weighted average transmittance over the relevant optical window, $\lambda$ denotes the wavelength, $T(\lambda, R_S)$ is the spectral transmittance evaluated at a fixed sheet resistance $R_S$, and $W(\lambda)$ is an application-dependent spectral weighting function, taken here as AM1.5G spectrum for photovoltaics, or the photopic response for displays (Fig. S1).

Obtaining $T_{app}$ for a given material involves the following steps:

(i) Selection of benchmark parameters (target $R_S$, optical window, Table 1) for the specific application.
(ii) Extraction of the transmission spectrum, $T(\lambda, R_S)$ at the corresponding thickness.
(iii) Integration of $T(\lambda, R_s)$ over the relevant optical window using $W(\lambda)$ to obtain $T_{app}$, as defined in Eq. 1.

Conceptually, the BEST framework enables performance evaluation under an equal-$R_s$ constraint, allowing direct comparison under application-relevant electrical conditions while avoiding the ambiguity of metrics that optimize performance by freely varying film thickness and sheet resistance. Expressed as $T_{app}$, the optical performance becomes a compact, application-specific quantity tied to the electrical requirements of the specific device category. We demonstrate this methodology for TCOs in photovoltaic and display contexts, although it is general and applicable to a wide range of optoelectronic system categories.

The BEST framework is rooted in device functionality, and thus its benchmarks are application-specific, in order to ensure its real-world relevance. We adopt representative benchmark parameters for PV and displays based on the literature (Table 1).[12–16] The PV benchmark represents crystalline-silicon electrodes, where photocurrents must spread laterally over mm–cm distances, necessitating low $R_s$ to limit electrical losses;[13] by contrast, the display benchmark reflects thin-film-transistor-controlled organic light-emitting diode (OLED) and liquid crystal display (LCD) electrodes, where localized pixel-level current delivery permits higher $R_s$, allowing better optical performance.[17] While the specific numerical ranges reflect representative implementations of these applications, the BEST framework is general and can be applied to alternative electrical constraints.

**Table 1.** Proposed BEST benchmark parameters for PV and display applications.[12–16]

| Application | Target $R_S$ (Ω/□) | Optical window (nm) |
|---|---|---|
| PV | 15-40 | 350-1200 |
| Displays | 100-150 | 380-780 |

To demonstrate the BEST framework, we analyze the common TCOs, indium tin oxide (ITO) and fluorine-doped tin oxide (FTO), as well as the emerging high-mobility indium-oxide-based TCOs: hydrogen-doped indium oxide (IO:H) and indium molybdenum oxide (IMO).[7,18–26] These materials represent established industrial standards and recently-developed TCOs, enabling a comparison between mature and emerging TCO paradigm. For each material, BEST maps provide a performance landscape versus wavelength and target sheet resistance, revealing the regions of applicability within the TCO design space. The construction of $T(\lambda, R_S)$ from the thickness-dependent transmittance and sheet-resistance relation is described in the Supplementary Information.

In these BEST maps, each material exhibits a distinct spectral dependence of the transparency-conductivity trade-off (Fig. 1). For ITO and FTO (Fig. 1a,b), transmittance remains high across much of the display window, but the iso-transmittance contours are strongly curved, indicating that the spectral response changes markedly as $R_S$ falls; in the PV window, transmittance then drops rapidly at low $R_S$, particularly toward the optical window edges. By contrast, IO:H and IMO (Fig. 1c,d) show nearly parallel contour lines over a broad range of $R_S$, implying a weaker $R_S$-dependence of $T(\lambda, R_S)$ and a substantially reduced long-wavelength penalty. The emerging TCOs therefore exhibit a more favorable transparency-conductivity balance, particularly in the long-wavelength regime relevant to PV.

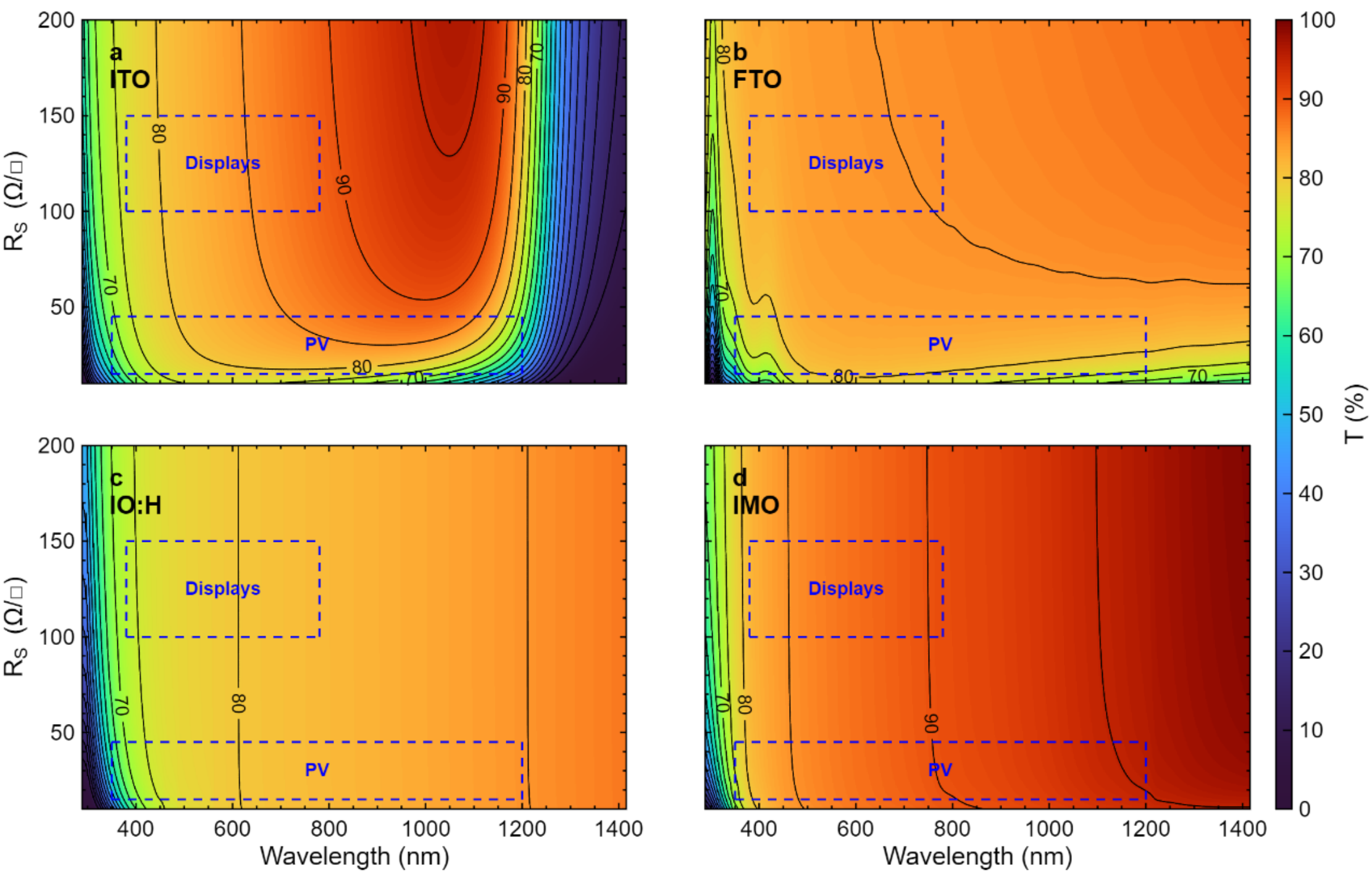

**Figure 1.** BEST transmittance maps. $T(\lambda, R_S)$ maps for (a) ITO, (b) FTO, (c) IO:H, and (d) IMO, evaluated at the thickness required to achieve each target $R_S$ from $R_S(t)$. Colors/contours indicate transmittance. Blue rectangles denote the PV and display benchmark windows (Table 1).

This map-level view also exposes a key limitation of current FOMs: when large high transmittance plateaus exist, the reported "optimal" value can be driven primarily by thickness-induced reductions in sheet resistance, rather than by intrinsic material performance. Since $R_S$decreases as film thickness increases, the near-parallel high-transmittance contours for IO:H and IMO mean that a FOM such as $T^{10}/R_S$ can be driven upward by moving toward thicker, lower-$R_S$ films, even when those films are no longer relevant as practical electrodes. Despite the richness of information in Fig. 1, it may be inconvenient for practical material selection. This motivates the reduction of the map representation into a metric that enables equal-$R_S$ comparison.

These two-dimensional maps provide a complete, constraint-resolved view of the transparency-conductivity trade-off, exposing how spectral performance evolves with $R_S$.

While the maps provide a complete view of the transparency–conductivity landscape for a given material, material selection requires comparison at specific operating points. To enable such comparison, we begin by interpreting the two-dimensional transmittance landscape $T(\lambda, R_S)$ within application-relevant operating windows. For a fixed $R_S$, each horizontal slice of the heat map (Fig. 1) corresponds to a transmittance spectrum representative of a realistic device operating point (Fig. 2a,b). These spectra directly illustrate wavelength-dependent trade-offs. For example, for PV, FTO outperforms IO:H at short wavelengths near the band edge, while IO:H outperforms at longer wavelengths, resulting in a spectral crossover (arrow, Fig. 2a). A similar wavelength-dependent reordering is observed in the display window, where FTO and ITO exchange ranking across the visible spectrum (arrow, Fig. 2b).

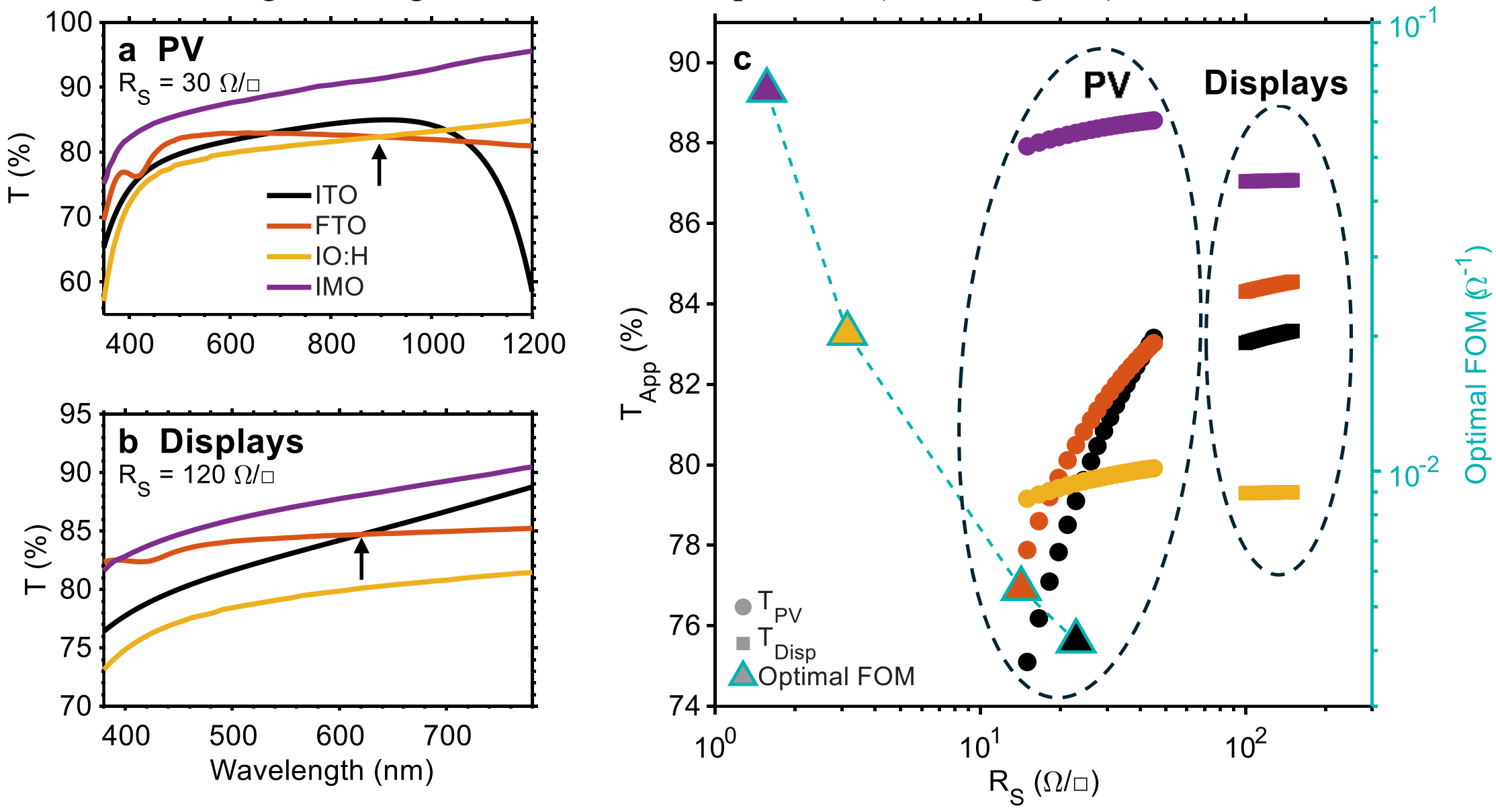

**Figure 2.** Spectral transmittance of ITO, FTO, IO:H, and IMO evaluated at representative sheet resistances for (a) photovoltaic ( $R_s = 30\ \Omega/\square$ ) and (b) display electrodes ( $R_s = 120\ \Omega/\square$ ). The arrows highlight wavelengths where the material ranking changes within the relevant spectral window. (c) $T_{app}$ as a function of $R_s$ for the photovoltaic ($R_s = 15 - 45\ \Omega/\square$, AM1.5G, circles) and displays ($R_s = 100 - 150\ \Omega/\square$, photopic response, squares) benchmark windows, obtained by applying Eq. (1). Triangles mark the operating points selected by maximizing the Haacke-type FOM. Dashed ellipses indicate the application-relevant PV and display $R_s$ ranges, while the dashed turquoise line connects the Haacke FOM-selected points. Complete FOM(t) curves are provided in the Supplementary Information (Fig. S5).

Applying Eq. (1) to these spectra yields the $T_{app}$ metric for a chosen $R_S$. Repeating this procedure across the relevant $R_S$ range constructs the $T_{app}(R_S)$ curve, which projects equal-$R_S$ optical performance across the application window (Fig. 2c). In this form, the curvature directly encodes the susceptibility of transparency to electrical constraint, revealing whether it is sensitive to $R_s$ or robust across the operating range. For example, ITO and FTO exhibit pronounced curvature within the PV range but remain comparatively stable across the display range. Moreover, the $T_{app}(R_S)$ curves enable direct, equal-$R_S$ comparison between materials and resolve wavelength-dependent crossovers inherent in the two-dimensional $T(\lambda, R_S)$ landscape. Importantly, through the modeled $R_S(t)$ relation (see Supporting Information), this representation links electrical constraint back to physical thickness (Fig. S5 and discussion therein), allowing to assess practical feasibility.

Haacke-type FOMs optimize thickness, effectively asking "*at what thickness does $T^{10}/R_S$ peaks?*" (Fig. S5). In contrast, $T_{app}(R_S)$ anchors the comparison to the electrical requirements of device operation, asking instead: "*at the sheet resistance my device requires, what transparency is obtained?*". Because these approaches optimize fundamentally different quantities, they can yield different conclusions. First, they can produce different material rankings: in the displays range, the Haacke-type FOM yields IO:H > FTO > ITO (Fig. 2c, squares), whereas $T_{App}(R_S)$ reorders this to FTO > ITO > IO:H (Fig. 2c, triangles). Second, these approaches can select different operating regimes: FOM optimization often corresponds to $R_S$ (and thickness) outside the regime targeted by real devices. In the display range, for example, the FOM maxima for all materials occurs at extremely low $R_S$ values that would rarely be pursued in pixel-level electrodes. These values correspond to very thick films, which are rarely adopted in practical device architectures due to fabrication, cost, and integration constraints.[12,16,27–29] The strong dependence of the Haacke metric on the $T^{10}$ term can amplify modest transparency gains into order of magnitude FOM improvements, creating large performance gaps which carry little practical relevance. By contrast, $T_{app}(R_s)$ provides a directly interpretable metric, enabling performance to be quantified as absolute transparency gains at a specified $R_S$ (e.g., a transparency increase of 3% between FTO and IMO at $R_s = 120\ \Omega/\square$, Fig. 2c).

While the BEST framework establishes a physically grounded basis for comparing transparent conductors under application-relevant electrical constraints, it is intentionally focused on isolating the intrinsic optical-electrical trade-off. In practice, material selection also depends on additional considerations, including availability, cost, and compatibility with large-scale processing. Integration constraints, such as deposition temperature, thermal and chemical stability, and interface quality, further influence whether a given material can be incorporated into a specific device architecture. Current injection between the TCO and device also plays a critical role in performance, which is not addressed here. These factors are inherently system-dependent and cannot be fully captured within a general benchmarking framework. Rather, the strength of BEST lies in providing a well-defined comparative baseline for transparent conductors under fixed electrical boundary conditions. By evaluating the TCO independently of a specific device stack, the framework isolates its intrinsic optical–electrical response and establishes a common reference point that can be used to guide material development, selection and subsequent integration into device architectures.

In conclusion, we introduce a framework for evaluating transparent conductors under fixed, application-relevant electrical constraints. By anchoring comparison to sheet resistance, performance is assessed at operating points relevant to the intended device operation. This approach yields two complementary representations: spectral maps $T(\lambda, R_S)$, which situate a given material within the transparency–conductivity landscape and reveal where it is best suited, and the reduced metric $T_{app}(R_S)$, which enables direct comparison between materials for a given application by quantifying absolute transparency gains at a specified sheet resistance. Applied to representative conventional and emerging TCOs, the framework reveals that material rankings are not intrinsic, but depend strongly on the sheet-resistance range of interest. More broadly, this work establishes a general strategy for evaluating materials under fixed operational constraints, providing a physically grounded basis for comparison and a transparent starting point for assessing compatibility with specific device architectures.

## References

1. Hosono, H. & Ueda, K. Transparent Conductive Oxides. in *Springer Handbook of Electronic and Photonic Materials* (eds Kasap, S. & Capper, P.) 1–1 (Springer International Publishing, Cham, 2017). doi:10.1007/978-3-319-48933-9_58.

2. Ginley, D., Hosono, H. & Paine, D. Handbook of Transparent Conductors Springer. *N. Y.* (2010).

3. Ellmer, K. Past achievements and future challenges in the development of optically transparent electrodes. *Nat. Photonics* **6**, 809–817 (2012).
4. Hosono, H. Recent progress in transparent oxide semiconductors: Materials and device application. *Thin Solid Films* **515**, 6000–6014 (2007).
5. Granqvist, C. G. Transparent conductors as solar energy materials: A panoramic review. *Sol. Energy Mater. Sol. Cells* **91**, 1529–1598 (2007).
6. Haacke, G. New figure of merit for transparent conductors. *J. Appl. Phys.* **47**, 4086–4089 (1976).
7. Zhang, L. *et al.* Correlated metals as transparent conductors. *Nat. Mater.* **15**, 204–210 (2016).
8. van Hest, M. F. A. M., Dabney, M. S., Perkins, J. D. & Ginley, D. S. High-mobility molybdenum doped indium oxide. *Thin Solid Films* **496**, 70–74 (2006).
9. Park, Y. *et al.* $SrNbO_3$ as a transparent conductor in the visible and ultraviolet spectra. *Commun. Phys.* **3**, 102 (2020).
10. Ma, C.-H. *et al.* Flexible transparent heteroepitaxial conducting oxide with mobility exceeding 100 $cm^2 V^{-1} s^{-1}$ at room temperature. *NPG Asia Mater.* **12**, 70 (2020).
11. Spurgeon, S. R. *et al.* Born-Qualified: An Autonomous Framework for Deploying Advanced Energy and Electronic Materials. Preprint at https://doi.org/10.48550/arXiv.2605.00639 (2026).
12. De Wolf, S., Descoeudres, A., Holman, Z. C. & Ballif, C. High-efficiency Silicon Heterojunction Solar Cells: A Review. *Green* **2**, 7–24 (2012).
13. Lin, H. *et al.* Silicon heterojunction solar cells with up to 26.81% efficiency achieved by electrically optimized nanocrystalline-silicon hole contact layers. *Nat. Energy* **8**, 789–799 (2023).

14. Yang, L. *et al.* Study and development of rear-emitter Si heterojunction solar cells and application of direct copper metallization. *Prog. Photovolt. Res. Appl.* **26**, 385–396 (2018).

15. Sheet Resistance Measurement. https://suragus.com/sheet-resistance/ (2025).

16. Morales-Masis, M., De Wolf, S., Woods-Robinson, R., Ager, J. W. & Ballif, C. Transparent Electrodes for Efficient Optoelectronics. *Adv. Electron. Mater.* **3**, 1600529 (2017).

17. He, Y., Hattori, R. & Kanicki, J. Four-Thin Film Transistor Pixel Electrode Circuits for Active-Matrix Organic Light-Emitting Displays. *Jpn. J. Appl. Phys.* **40**, 1199 (2001).

18. Minenkov, A. *et al.* Monitoring the Electrochemical Failure of Indium Tin Oxide Electrodes via Operando Ellipsometry Complemented by Electron Microscopy and Spectroscopy. *ACS Appl. Mater. Interfaces* **16**, 9517–9531 (2024).

19. Asl, H. Z. & Rozati, S. M. High-quality spray-deposited fluorine-doped tin oxide: effect of film thickness on structural, morphological, electrical, and optical properties. *Appl. Phys. A* **125**, 689 (2019).

20. Ching-Prado, E., Watson, A. & Miranda, H. Optical and electrical properties of fluorine doped tin oxide thin film. *J. Mater. Sci. Mater. Electron.* **29**, 15299–15306 (2018).

21. Budhi, A. W. S. Hydrogenated indium oxide (IO: H) for thin film solar cell. (Master's thesis, Delft University of Technology, 2016).

22. Koida, T., Fujiwara, H. & Kondo, M. Hydrogen-doped In2O3 as high-mobility transparent conductive oxide. *Jpn. J. Appl. Phys.* **46**, L685 (2007).

23. Egbo, K. O., Adesina, A. E., Ezeh, C. V., Liu, C. P. & Yu, K. M. Effects of free carriers on the optical properties of high mobility transition metal doped $In_2O_3$ transparent conductors. *Phys. Rev. Mater.* **5**, 094603 (2021).

24. Gupta, R. K., Ghosh, K., Patel, R. & Kahol, P. K. Effect of thickness on optoelectrical properties of Mo-doped indium oxide films. *Appl. Surf. Sci.* **255**, 3046–3048 (2008).

25. Chen, H.-Y. *et al.* Indium-doped molybdenum oxide as a new p-type transparent conductive oxide. *J. Mater. Chem.* **21**, 5745–5752 (2011).

26. Meng, Y. *et al.* Molybdenum-doped indium oxide transparent conductive thin films. *J. Vac. Sci. Technol. Vac. Surf. Films* **20**, 288–290 (2002).

27. A. Jacobs, D., R. Catchpole, K., J. Beck, F. & P. White, T. A re-evaluation of transparent conductor requirements for thin-film solar cells. *J. Mater. Chem. A* **4**, 4490–4496 (2016).

28. Transparent Conductive Oxides: Implications for Industry Standards. *PatSnap Eureka* https://eureka.patsnap.com/report-transparent-conductive-oxides-implications-for-industry-standards.

29. Gordon, R. G. Technological Challenges for Transparent Conductors. *Adv. Sci. Technol.* **33**, 1037–1050 (2003).

# SUPPLEMENTARY INFORMATION

# Benchmarking Transparent Conductors

**Amit Cohen, Lior Kornblum**

Andrew and Erna Viterbi Department of Electrical and Computer Engineering,

Technion—Israel Institute of Technology, Haifa 32000-03, Israel

## 1. Construction of the Thickness-dependent Transmittance

For each material, the wavelength and thickness-dependent transmittance $T(\lambda, t)$,[1] was constructed from the complex refractive index of the TCO film, written as

$$N(\lambda) = n(\lambda) + ik(\lambda),$$

where $n$ is the real refractive index and $k_f$ is the extinction coefficient. The absorption coefficient is then given by

$$\alpha(\lambda) = \frac{4\pi k(\lambda)}{\lambda},$$

where $\lambda$ and $t$ are expressed in consistent units. To avoid introducing substrate-specific optical effects, the film transmittance was calculated for a free-standing air/TCO/air geometry at normal incidence. In this approximation, the film is treated as an absorbing layer surrounded by air on both sides. The intensity reflectance at the air/TCO interface is

$$R(\lambda) = \left|\frac{1 - N(\lambda)}{1 + N(\lambda)}\right|^2.$$

Because the surrounding medium is air on both sides, the reflectance at the two film interfaces is the same. The single-pass absorption factor through a film of thickness $t$ is

$$A(\lambda, t) = e^{-\alpha(\lambda)t}.$$

The total incoherent transmittance through the air/TCO/air structure is then obtained by summing the multiple internally reflected intensity contributions:

$$T(\lambda, t) = \frac{\left(1 - R(\lambda)\right)^2 e^{-\alpha(\lambda)t}}{1 - R^2(\lambda)e^{-2\alpha(\lambda)t}}.$$

This expression represents the fraction of incident light that enters the TCO film, survives absorption during propagation through the film, and exits into air, including the contribution from incoherent multiple internal reflections.

## 2. Spectral Weighting Functions

The spectral weighting functions used to calculate $T_{app}(R_s)$ are shown in Fig. S1. The photovoltaic weighting was based on the AM1.5G solar spectrum, while the display weighting was based on the CIE photopic response.[2,3] These functions define the wavelength regions that contribute most strongly to the application-weighted transmittance values discussed in the main text.

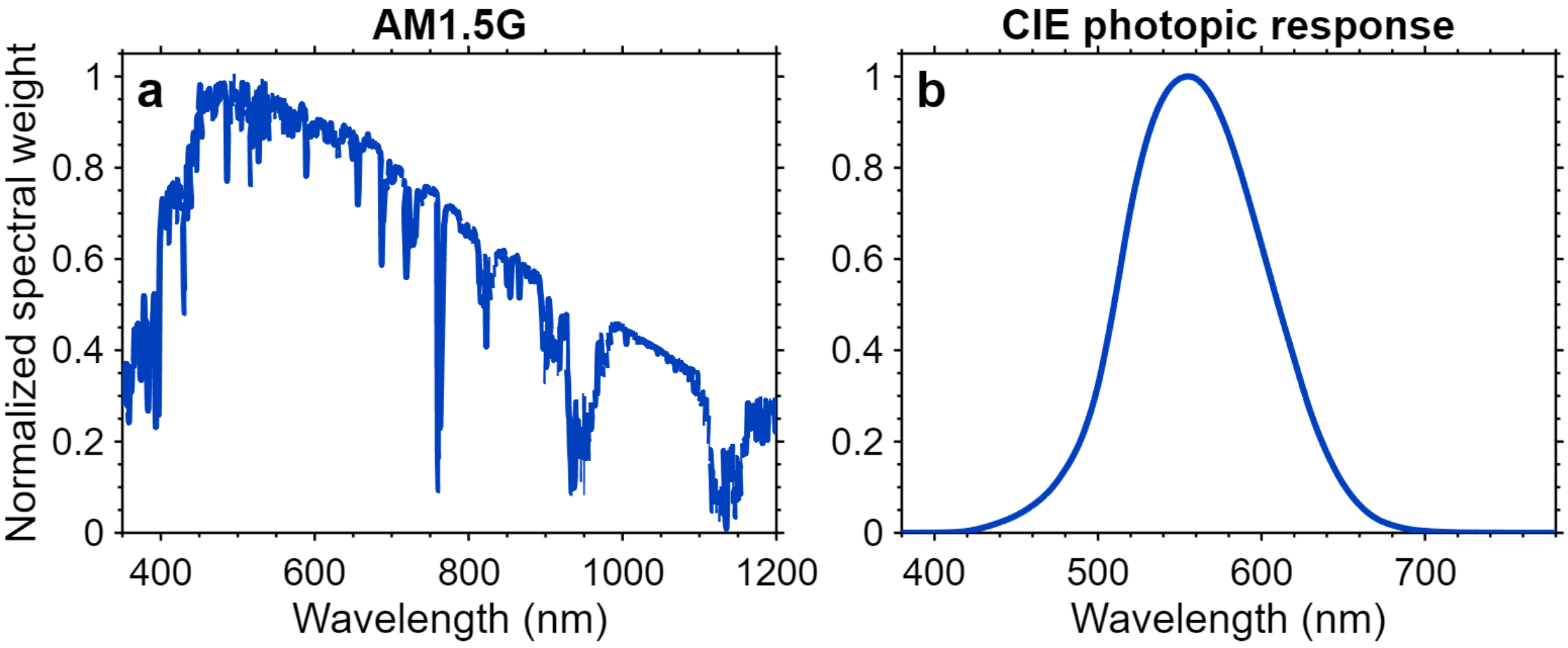

**Figure S1**. Spectral weighting functions used for application-weighted BEST analysis. Normalized (a) AM1.5G solar spectrum and (b) CIE photopic response function used for the photovoltaic- and display-relevant transmittance calculations, respectively. Both curves are normalized to their maximum value for visual comparison.

## 3. Sheet Resistance Modeling

The electrical performance is strongly thickness-dependent, therefore, establishing a physically motivated reliable $R_s(t)$ relation across a realistic range is essential for device design and benchmarking.

At large thicknesses ($t \gg \Lambda$), where $\Lambda$ is the electron mean-free path (EMFP), transport is bulk-like and described simply by $R_s(t) = \rho_0 / t$ where $\rho_0$ is the bulk resistivity (Fig. S2). This limit where $\rho$ is constant sets the benchmark for intrinsic material performance. For thin films, additional scattering effects become important, and more advanced models are required. When film thickness approaches EMFP ($t > \Lambda$), surface scattering must be considered. In this regime, grain size often also needs to be taken into account. The Fuchs–Sondheimer (FS) model[4,5] accounts for diffuse versus specular surface scattering, while the Mayadas–Shatzkes (MS) model[6] treats grain boundary reflection. In practice, these are often combined to describe thin film transport realistically. For ultrathin films ($t \sim \Lambda$), just above the percolation threshold, the film behaves as a composite of metallic regions and voids. While not universally accurate, effective medium approximation (EMA) combined with Drude conductivity can model this regime, and sometimes it provides reasonable scaling in the transitional thickness range before FS/MS behavior dominates.[7] In the extreme ultrathin limit of only a few monolayers ($t < \Lambda$), the film may no longer remain fully continuous, and electrical transport becomes dominated by tunneling between isolated metallic islands, conditions under which classical Rs(t) modeling is no longer valid.[8] Even in the case of continuous epitaxial films, quantum size effects (QSEs) can emerge due to electron confinement.[9] These regimes lie well outside the practical thickness range typically considered for TCO device applications.[10–12]

Within the thickness range relevant to the present TCO analysis, the films are treated as continuous conducting layers. Accordingly, deviations from the bulk $R_s = \rho_0/t$limit are described using a finite-size-scattering model, consistent with the thickness range of the literature datasets used here. We therefore model $R_s(t)$ using a combined Fuchs–Sondheimer/Mayadas–Shatzkes approach, as described in the following section. This goal is pragmatic as well as physical: to generate a physically constrained, monotonic $R_s(t)$ relation that can be inverted to obtain $t(R_S)$. This relation is then used to construct the BEST maps, $T(\lambda; R_S)$, under fixed sheet-resistance constraints.

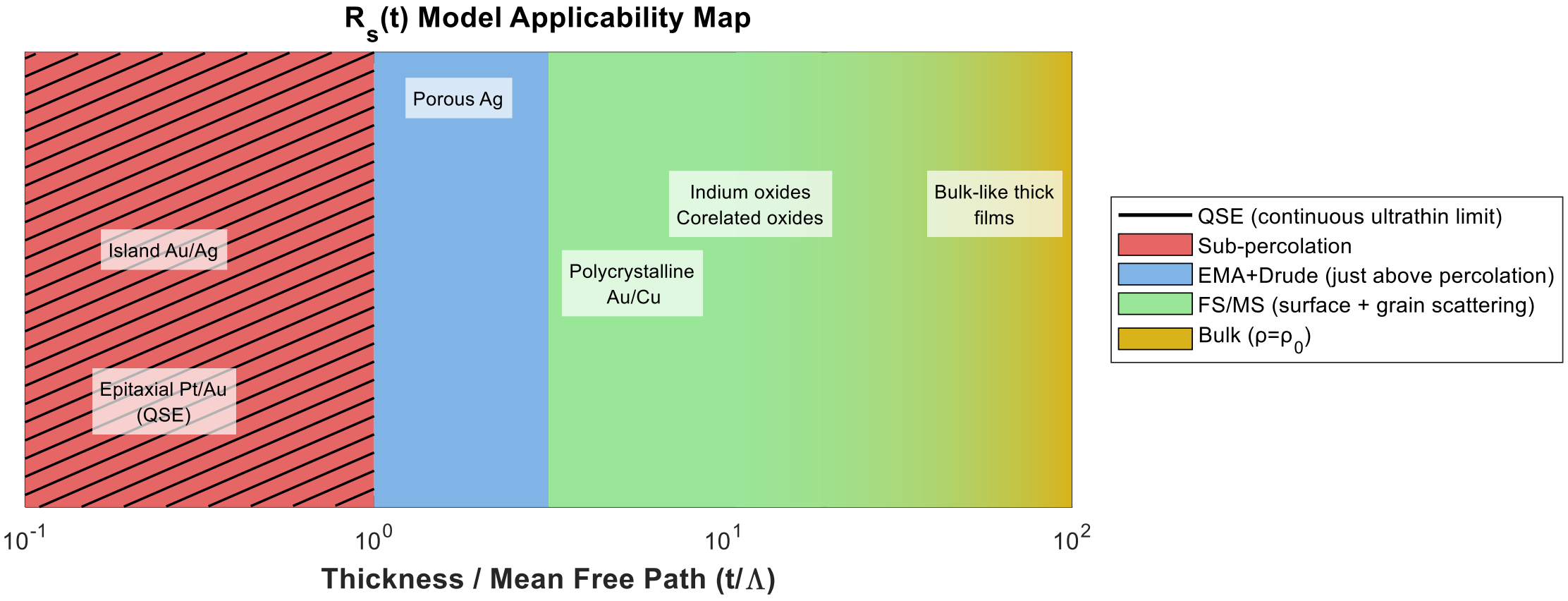

Figure S2. Applicability map of Rs(t) models as a function of normalized thickness ($t/\Lambda$), where $\Lambda$ is the EMFP. As the film becomes thinner, transport evolves from bulk-like conduction ($\rho \approx \rho_0$) through classical finite-size scattering described by the FS/MS models, to a near-percolation regime that can be approximated by an EMA approach. Below the percolation threshold, the film becomes discontinuous and cannot be described by classical Rs(t) models. In atomically continuous ultrathin films, (QSE) may arise due to electron confinement. Representative material examples are indicated within each regime.

# 4. Construction of the Thickness-dependent Sheet Resistance

The measured sheet resistance data were fitted using a combined finite-size-scattering model. The sheet resistance was calculated as

$$R_S(t) = \frac{\rho(t)}{t},$$

The thickness-dependent resistivity is written as

$$\rho(t) = \rho_{FS}(t) + \rho_{MS} - \rho_0,$$

with the FS resistivity contribution

$$\rho_{FS}(t) = \frac{\rho_0}{S_{FS}(t/\Lambda, p)}$$

and the MS conductivity ratio is written as

$$\rho_{MS} = \frac{\rho_0}{S_{MS}(\xi)}.$$

The FS conductivity ratio is evaluated as

$$S_{FS}(\kappa, p) = 1 - \frac{3}{2}(1-p)\int_0^1 \mu(1-\mu^2)\frac{1-e^{-\frac{\kappa}{\mu}}}{1-pe^{-\frac{\kappa}{\mu}}}d\mu$$

where

$$\kappa = \frac{t}{\Lambda},$$

Here, p is the surface specularity parameter. The limit $p = 1$ corresponds to fully specular surface scattering, for which the FS correction vanishes. Lower values of prepresent more diffuse surface scattering and therefore stronger finite-size enhancement of the resistivity.

The MS conductivity ratio was written as

$$S_{MS}(\xi) = 1 - \frac{3}{2}\xi + 3\xi^2 - 3\xi^3 \ln\left(1+\frac{1}{\xi}\right).$$

The dimensionless parameter $\xi$ represents the effective grain-boundary scattering strength,

$$\xi = \frac{\Lambda}{D}\frac{R}{1-R},$$

where D is the characteristic grain size and R is the grain-boundary reflection coefficient. In this work, $\xi$ was treated as an effective fitting parameter, because the literature datasets used for the BEST construction do not provide a consistent independent determination of both D and R for all samples.

Therefore, the fitted expression is

$$R_s(t) = \frac{1}{t}\left[\frac{\rho_0}{S_{FS}(t/\Lambda, p)} + \frac{\rho_0}{S_{MS}(\xi)} - \rho_0\right].$$

The subtraction of $\rho_0$ avoids double-counting the bulk resistivity contribution, since both the FS and MS terms reduce to $\rho_0$ in the absence of additional scattering. The resulting fits and parameters are shown in Figs. S3,4 and Table S1, respectively.

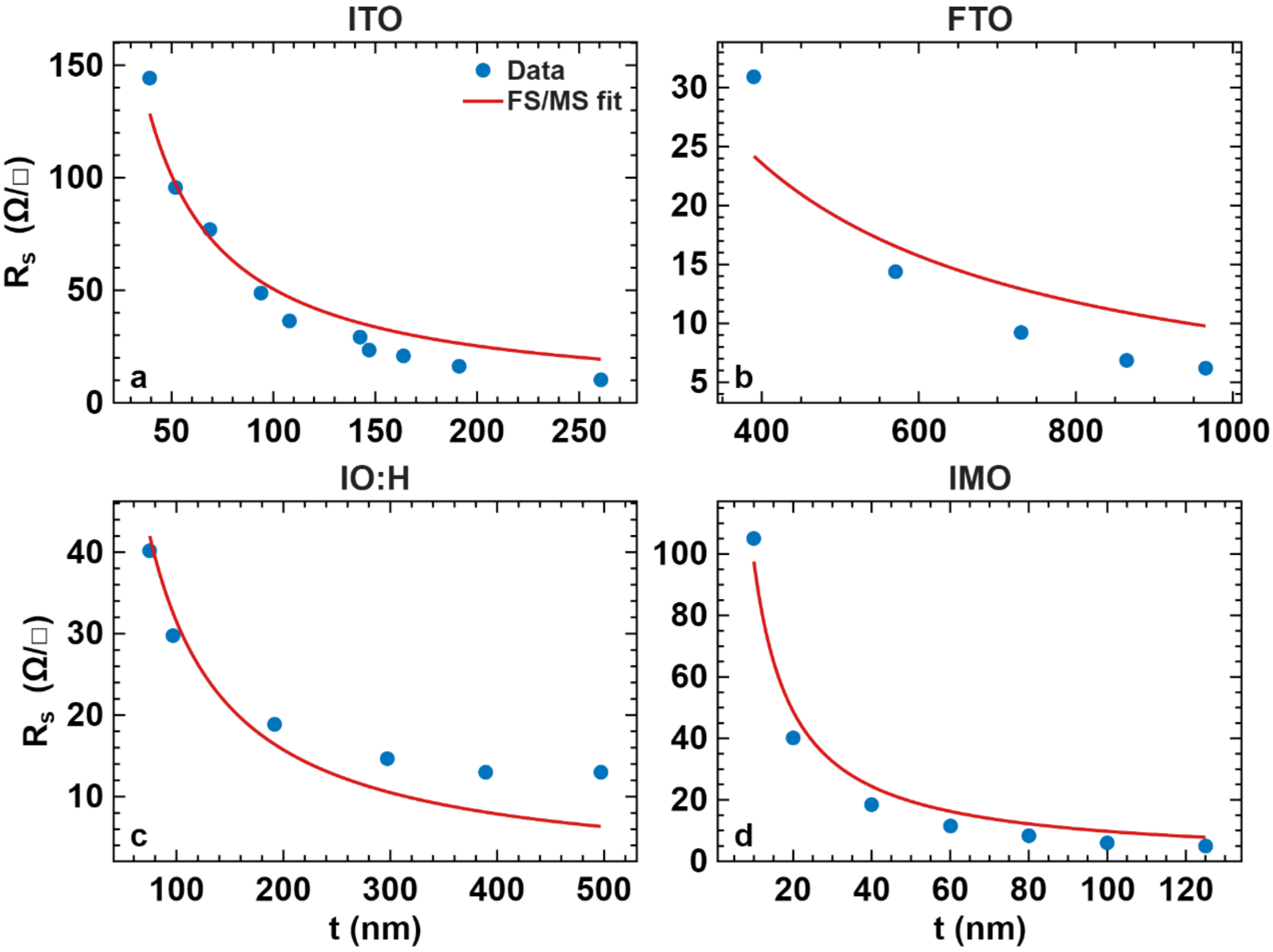

**Figure S3.** Thickness-dependent sheet-resistance fits used for BEST-map construction for (a) ITO, (b)

FTO, (c) IO:H, and (d) IMO.[13–19] The red curves show the fitted finite-size-scattering model combining FS surface scattering and MS grain-boundary scattering.

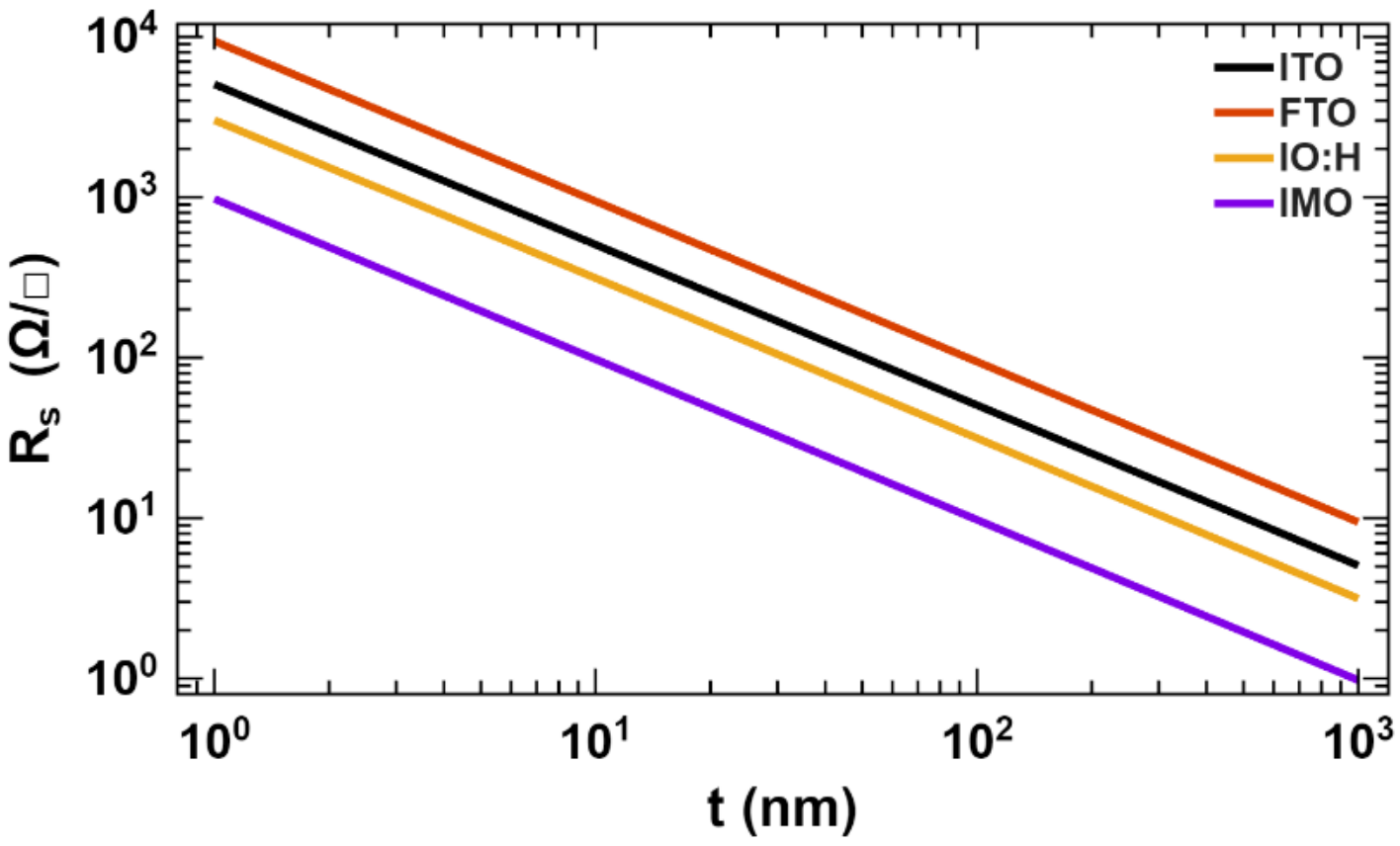

**Figure S4**. Modeled thickness-dependent sheet resistance for all TCOs. Modeled $R_S(t)$ curves for ITO, FTO, IO:H, and IMO plotted on a log–log scale, corresponding to the individual fits shown in Fig. S6.

**Table S1.** FS/MS fit parameters for all TCO samples. $\rho_0$ is the bulk resistivity, Λ is the EMFP calculated from $\rho_0$ and the charge carrier density (n), p is the FS surface specularity (0 = fully diffuse, 1 = fully specular), and $\xi = (\Lambda/D) \cdot R/(1-R)$ is the MS grain-boundary scattering parameter. $\rho_0$ and n were taken from the literature, while p and ξ were fitted. ).[13–19]

| Material | $\rho_0$ (Ω · cm) | n ($cm^{-3}$) | Λ (nm) | p | ξ |
|---|---|---|---|---|---|
| ITO | $2.70\times10^{-4}$ | $4.50\times10^{20}$ | 8.0 | 1.00 | 0.63 |
| FTO | $6.00\times10^{-4}$ | $2.60\times10^{20}$ | 3.70 | 0.87 | 0.37 |
| IO:H | $2.60\times10^{-4}$ | $1.50\times10^{20}$ | 17.3 | 0.80 | 0.09 |
| IMO | $6.07\times10^{-5}$ | $3.92\times10^{20}$ | 39.1 | 1.00 | 0.43 |

## 5. Thickness-dependent FOM Analysis

The Haacke-type FOM was evaluated explicitly as a function of film thickness, illustrating how thickness optimization differs from fixed-$R_S$ benchmarking (Fig. S5). The FOM is given by $\phi = T(t)^{10}/R_S(t)$, where T is the spectrally weighted average transmittances t, chosen here to allow direct comparison with $T_{app}$.

Because the spectral weighting depends on the application, two FOMs can be defined: $\phi_{PV}$, calculated using the AM1.5G weighting, and $\phi_{Disp}$, calculated using the CIE photopic weighting. These are formally distinct quantities and therefore yield separate optimal $R_S$ values and film thicknesses. For a given material, the optimal $R_S$ values obtained from the PV- and display-weighted FOMs are very similar (Fig. S5a,b). For example, the FOM optimum for IMO occurs at $R_S \approx 2\ \Omega/\square$ using the PV weighting and $R_S \approx 1\ \Omega/\square$ using the display weighting. These differences are small compared with the much larger separation between the optimal FOM $R_S$ values and the application-relevant windows (Fig. 2c). For clarity, the main text therefore shows a single representative optimal-FOM marker for both PV and display comparisons.

The difference between the PV- and display-optimized thicknesses can be more pronounced than the difference in optimal $R_S$. Nevertheless, the conclusion remains unchanged: except for cases such as ITO and FTO under PV weighting, where the FOM optimum falls near the PV-relevant $R_S$ range, the FOM-selected thicknesses generally lie far from the thicknesses associated with application-specific windows.

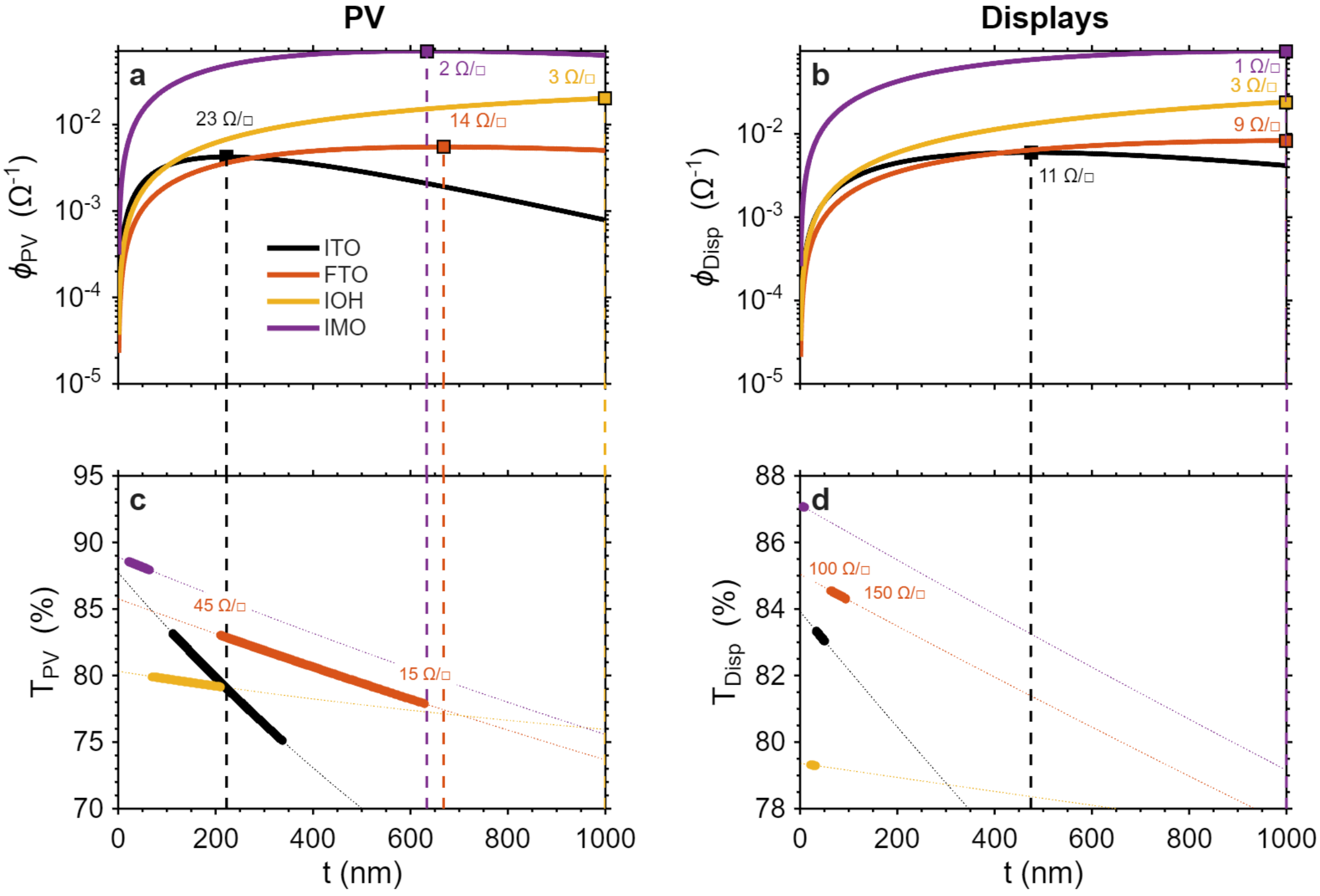

**Figure S5.** Thickness-dependent Haacke-type FOM and $T_{app}$. (a,b) Thickness-dependent Haacke-type FOMs calculated using PV and display spectral weightings, respectively. Square markers indicate the FOM maxima, and dashed vertical lines indicate the corresponding thicknesses. (c,d) Corresponding $T_{app}$ plotted as functions of thickness. The thick solid segments indicate the thickness ranges corresponding to the application-relevant $R_S$windows used in the BEST analysis, as defined in Table 1. The dotted extensions show $T_{app}$ outside these $R_S$ windows and are included only to illustrate how the FOM-selected thicknesses compare with the application-constrained regions.

## References

1. Swanepoel, R. Determination of the thickness and optical constants of amorphous silicon. *J. Phys. [E]* **16**, 1214–1222 (1983).

2. Gueymard, C. A., Myers, D. & Emery, K. Proposed reference irradiance spectra for solar energy systems testing. *Sol. Energy* **73**, 443–467 (2002).

3. International Commission On Illumination (Cie). CIE spectral luminous efficiency for photopic vision. International Commission on Illumination (CIE) https://doi.org/10.25039/CIE.DS.dktna2s3.

4. Fuchs, K. The conductivity of thin metallic films according to the electron theory of metals. *Math. Proc. Camb. Philos. Soc.* **34**, 100–108 (1938).

5. Sondheimer, E. H. The mean free path of electrons in metals. *Adv. Phys.* **50**, 499–537 (2001).
6. Mayadas, A. F. & Shatzkes, M. Electrical-Resistivity Model for Polycrystalline Films: the Case of Arbitrary Reflection at External Surfaces. *Phys. Rev. B* **1**, 1382–1389 (1970).
7. Hövel, M., Gompf, B. & Dressel, M. Dielectric properties of ultrathin metal films around the percolation threshold. *Phys. Rev. B* **81**, 035402 (2010).
8. King, P. D. C. & Veal, T. D. Conductivity in transparent oxide semiconductors. *J. Phys. Condens. Matter* **23**, 334214 (2011).
9. Aballe, L., Rogero, C. & Horn, K. Quantum size effects in ultrathin epitaxial Mg films on Si(111). *Phys. Rev. B* **65**, 125319 (2002).
10. Granqvist, C. G. Transparent conductors as solar energy materials: A panoramic review. *Sol. Energy Mater. Sol. Cells* **91**, 1529–1598 (2007).
11. Minami, T. Transparent conducting oxide semiconductors for transparent electrodes. *Semicond. Sci. Technol.* **20**, S35 (2005).
12. Hosono, H. Ionic amorphous oxide semiconductors: Material design, carrier transport, and device application. *J. Non-Cryst. Solids* **352**, 851–858 (2006).
13. Zhang, L. *et al.* Correlated metals as transparent conductors. *Nat. Mater.* **15**, 204–210 (2016).
14. Minenkov, A. *et al.* Monitoring the Electrochemical Failure of Indium Tin Oxide Electrodes via Operando Ellipsometry Complemented by Electron Microscopy and Spectroscopy. *ACS Appl. Mater. Interfaces* **16**, 9517–9531 (2024).
15. Asl, H. Z. & Rozati, S. M. High-quality spray-deposited fluorine-doped tin oxide: effect of film thickness on structural, morphological, electrical, and optical properties. *Appl. Phys. A* **125**, 689 (2019).

16. Ching-Prado, E., Watson, A. & Miranda, H. Optical and electrical properties of fluorine doped tin oxide thin film. *J. Mater. Sci. Mater. Electron.* **29**, 15299–15306 (2018).

17. Budhi, A. W. S. Hydrogenated indium oxide (IO: H) for thin film solar cell. (Master's thesis, Delft University of Technology, 2016).

18. Egbo, K. O., Adesina, A. E., Ezeh, C. V., Liu, C. P. & Yu, K. M. Effects of free carriers on the optical properties of high mobility transition metal doped $In_2O_3$ transparent conductors. *Phys. Rev. Mater.* **5**, 094603 (2021).

19. Gupta, R. K., Ghosh, K., Patel, R. & Kahol, P. K. Effect of thickness on optoelectrical properties of Mo-doped indium oxide films. *Appl. Surf. Sci.* **255**, 3046–3048 (2008).